\documentstyle[12pt]{article}
\begin{document}

\begin{center}
{\Large {\bf Impulsive spherical gravitational waves}}  \\[10mm]
{\Large A. N. Aliev and Y. Nutku}\\[4mm] Feza G\"ursey Institute, P. K. 6
\c{C}engelk\"oy 81220 Istanbul, Turkey\\[10mm]
November 6, 2000\\[10mm]
\end{center}

\noindent
  Penrose's identification with warp provides the general framework for
constructing the continuous form of impulsive gravitational wave metrics.
We present the $2$-component spinor formalism for the derivation of the full
family of impulsive spherical gravitational wave metrics which brings out
the power in identification with warp and leads to the simplest derivation
of exact solutions. These solutions of the Einstein vacuum field equations
are obtained by cutting Minkowski space into two pieces along a null cone
and re-identifying them with warp which is given by an arbitrary
non-linear holomorphic transformation. Using $2$-component spinor
techniques we construct a new metric describing an impulsive
spherical gravitational wave where the vertex of the null cone lies on a
world-line with constant acceleration.

\section{Introduction}

  For weak fields and slow motion, the emission of gravitational radiation
requires that the source must posses at least a quadrupole moment
\cite{einstein}. The experimental verification of the ``quadrupole formula"
has been spectacular \cite{taylor}. However, this radiation is weak, down by
a factor of $G/c^5$ which is an infamous fourty orders of magnitude.
On the other hand, an extrapolation of the quadrupole formula to
strong fields by dimensional analysis \cite{mtw} turns this factor around
to its inverse and suggests that gravitational radiation is the
mechanism for releasing most energy by anything anyhow.

    For strong gravitational fields there is a mechanism for the emission
of gravitational waves which makes no reference to quadrupole radiation.
It starts with an apparent paradox: Monopole radiation is forbidden by the
principle of equivalence, yet we have exact solutions of the Einstein field
equations that describe spherical gravitational waves.
These are the famous solutions of
Robinson and Trautman \cite{rt1}. Soon after their discovery it was realized
that Petrov Type $N$ Robinson-Trautman solutions posses ``wire singularities,"
that is, the wave-fronts are locally spherical but they are not complete.
If we cut a paraboloid out of Minkowski space, impose perfectly
reflecting boundary conditions and shine an impulsive
plane gravitational wave \cite{rp1} on it, then
an impulsive spherical wave will form in the neighborhood of the focus.
The paraboloid must be of finite extent for the spacetime to support the
impulsive plane wave and therefore a cone with the opening angle of the
paraboloid will be missing out of the wave-front.
Thus a contradiction with the principle of equivalence is avoided
but such a construction sheds little light on
the physical origin of Type $N$ Robinson-Trautman metrics.

Considerations of topological defects which may have formed
in the early universe \cite{kibble} provide an interesting
possibility for the physical interpretation of wire singularities
in radiative Robinson-Trautman metrics. Indeed, cosmic strings
which are physically the most feasible topological defects have
proved to be very popular for restoring to physics all sorts of
metrics with conical singularities. However, an examination of the
properties of continuous metric for an impulsive spherical wave
(\ref{harika}) shows that nowhere is the case for cosmic strings as
compelling as it is for impulsive spherical gravitational waves where
they have an unambigious role to play: {\it a snapping cosmic string
emits impulsive spherical gravitational waves and is thereby annihilated}.

The exact solution describing an impulsive spherical gravitational
wave cannot be given in the Robinson-Trautman form because of severe
discontinuities encountered in the Robinson-Trautman coordinate patch.
The continuous form of the metric is obtained by
the ``scissors and paste" approach that Penrose \cite{battelle}, \cite{bible}
has devised for the study of impulsive gravitational waves.
Impulsive wave metrics are obtained by cutting Minkowski space into
two pieces along a null hypersurface and gluing them back together
after warp. For impulsive spherical gravitational waves the null
hypersurface is a cone and Penrose's identification with warp yields
the metric \cite{np}
\begin{equation}
 d s^{2} = 2\,d \,u \,d \, v  - 2 \left| u \, d \, {\bar \zeta} +
  v \, \theta(v) \{ h \, ; \zeta \}\,d \,\zeta \right|^2
\label{harika}
\end{equation}
where $\theta$ is the Heaviside unit step function, $h$ is
an arbitrary holomorphic function
\begin{equation}
 h_{\bar\zeta}=0
\end{equation}
which determines the warp and
\begin{equation}
   \{ h \, ; \zeta \} = - \frac{1}{2} \left(
   \frac{h'''}{h'}
 - \frac{3}{2} \; \frac{{h''}^{\;2}}{{h'}^{\;2}}
 \right)                       \label{schwarz}
\end{equation}
is its Schwarzian derivative. This metric satisfies Einstein's vacuum
field equations for all $v$ and it is continuous across the null hypersurface
$v=0$. It is Petrov Type $N$ with the Weyl curvature suffering a Dirac
$\delta$-function discontinuity at $v=0$ that establishes the impulsive
character of the spherical wave. The metric (\ref{harika}) remained
unpublished until 1992 even though quantum effects in its background were
studied by Horta\c{c}su \cite{hortacsu} which marked its first appearance in
print. Meanwhile Gleiser and Pullin \cite{gp} have independently found a
special case of eq.(\ref{harika}) which results from the choice of an
exponential for the warp function and gave its correct interpretation as the
radiation accompanying the snapping of a straight, infinitely long cosmic
string. Later, Hogan has repeated the derivation of the
metric (\ref{harika}) for an arbitrary holomorphic warp function.
Griffiths and Podolsk\'y \cite{gp1}, \cite{gp2} have presented exact
solutions for impulsive spherical waves in de Sitter and anti-de Sitter
spacetimes. They have also discussed the relationship of impulsive spherical
wave solutions to collision and snapping of cosmic strings \cite{gp3}.
A different approach to the collision problem was proposed by Tod \cite{tod}.
We shall not discuss the collision problem. In our case the physical
interpretation of the impulsive spherical wave metrics requires a
cosmic string that has already snapped and is acting as the source
of the radiation. This is based on the fact that Minkowski
space outside the light cone has a conical deficit whereas inside it is
complete.

In this paper we shall present $2$-component spinor techniques to
construct an arbitrary holomorphic transformation of Minkowski
space which is the proper framework for constructing impulsive spherical
wave metrics. This is a point emphasized by Penrose \cite{bible}.
Penrose's identification with warp imposes continuity conditions
on the spinor field defining the null frame on both sides of
Minkowski space after the removal of the null cone. We shall also
discuss physical properties of this metric and show that in the
exterior region of the future null cone $v>0$ the metric exhibits
conical singularities characteristic of a cosmic string.
This means that the two regions of Minkowski space $v<0$
and  $v>0$ are matched at $v=0$ and the full spacetime obtained by
the union of these three pieces describes a cosmic string which
has already snapped and its free ends are moving away at the speed of light.
The impulsive spherical gravitational wave is at the lightcone
and flat spacetime without conical deficit is left behind. Thus the
cosmic string is getting annihilated as
the impulsive spherical gravitational wave propagates.

    We shall present a new Type $N$ exact solution of the Einstein field
equations describing an impulsive spherical gravitational wave that
incorporates uniform acceleration into the solution. Petrov Type $N$
Robinson-Trautman spherical
gravitational wave metrics contain a parameter $k=0,\,\pm1 $ which
can be regarded as specifying the world-line of a corresponding source
for spherical gravitational waves and in identification with warp it
coincides with
the world-line of the vertex of the null cone. The metric (\ref{harika})
corresponds to the null, $k=0$, case. We shall show that it is also possible
to choose a world-line subject to constant acceleration. This is
possible only for non-vanishing $k$. We shall therefore start by
applying Penrose's identification with warp to obtain the impulsive
spherical wave metric for all values of the parameter $k$ which reduces
to Hogan's solution \cite{hogan1} for $k=+1$. Then we shall
present the new solution by constructing the arbitrary holomorphic
transformation of Minkowski space
for the case of a constant acceleration world-line.

\section{Penrose's identification with warp}

We recall \cite{bible} that the construction of an impulsive spherical
wave through Penrose's scissors and paste approach starts with the
removal of a null cone ${\cal N}$ from Minkowski space.
This leaves two disjoint Minkowskian regions and the metrics appropriate to
these regions are derived from the standard Minkowski interval
\begin{equation}
  ds^{2} =  2\,d\,U' d\,V' - 2\,d\,Z' d\,\bar{Z'}
\label{minkowski}
\end{equation}
by means of the transformation
\begin{eqnarray}
  U'&=& u, \nonumber \\
  V'&=& v + u\,|\zeta|^2, \label{trans}\\
  Z'&=& u\,\zeta  \nonumber
\end{eqnarray}
so that the metric of Minkowskian region ${\bf M}$ outside a future
null cone ${\cal N}$ $\,(v>0)\,$ is given by
\begin{equation}
  ds^{2} = \, 2\,d\,u \;d\,v \; - \; 2\,u^{2}\;d\,\zeta \;d\,\bar{\zeta},
\label{cone1}
\end{equation}
while another Minkowskian region ${\hat {\bf M}}$  inside
the cone ${\cal N}$ $\,(v<0)\,$ is described by the metric
\begin{equation}
  d{\hat s}^{2} = 2\,d\,{\hat u} \;d\,{\hat v} \;
  - \; 2\,{\hat u}^{2}\;d\,{\hat \zeta } \;d\,{\hat {\bar \zeta} }.
\label{cone}
\end{equation}
Here $\,v\,$ is a null coordinate,  $\, u\, $ is a Bondi-type
luminosity distance and $\,\zeta\,$ is the stereographic coordinate
on the Riemann sphere.
In Penrose's identification with warp these two disjoint
regions are reattached along a null cone ${\cal N}$
according to the relations
\begin{eqnarray}
 {\hat v} &=&0 \; = \; v, \nonumber \\[1mm]
 {\hat u} &=& \frac{u}{|h'|},    \label{warp} \\[1mm]
  {\hat \zeta} &=& h(\zeta)  \nonumber
\end{eqnarray}
where we recall that $h$ is an arbitrary holomorphic function
and prime denotes derivative with respect to its argument.
The $(u,\zeta)$-transformation in
eqs.(\ref{warp}) is an arbitrary holomorphic transformation
of the spin-space at the vertex of a future null cone ${\cal N}$.
In order to show this we choose at each point of the null cone ${\cal N}$
a $2$-component spinor $ \,\xi^A \,$ such that the position vector
of a point on the cone can be expressed as
$$ x^{\mu} \leftrightarrow x^{AX'} = u\,\xi^A  \, \xi^{X'} $$
where in terms of the coordinates (\ref{trans}) we can express this
by the trace-free spin matrix
\begin{equation}  \begin{array}{c}
\left(  \begin{array}{cc}
1 & {\bar \zeta} \\
\zeta  & \zeta  {\bar \zeta} \end{array} \right) =
\left(  \begin{array}{cc}
\xi^0  \, {\bar \xi}^{0'}      &   \xi^0  \, {\bar \xi}^{1'} \\
\xi^1  \, {\bar \xi}^{0'}      &   \xi^1  \, {\bar \xi}^{1'}
\end{array} \right)       \end{array}
\label{spinmat}
\end{equation}
and the $2$-component spinor is given by
\begin{equation}
 \xi^A=
 \left(  \begin{array}{c}
1  \\
\zeta \end{array} \right).
\label{2spinor}
\end{equation}
Penrose's requirement of Type $I$ geometry is expressed by the
invariance of the $1$-form
\begin{equation}
\alpha = \xi_{A} \, d \xi^{A}
\label{typeIg}
\end{equation}
under identification with warp. Explicitly,
an arbitrary holomorphic mapping $\zeta \rightarrow h(\zeta) $
preserving the differential form (\ref{typeIg}) on
the future null cone ${\cal N}$ corresponds to the $2$-component spinor
of the form
\begin{equation}
 \hat \xi^A=
\left(  \begin{array}{c}
h  \\ \zeta \,  \eta^{1/2} \end{array} \right)
\label{hspinor}
\end{equation}
where
\begin{equation}
\eta = \frac{1}{h'}
\label{etatr}
\end{equation}
which yields the last one of eqs.(\ref{warp}) in the identification with
warp. The Hamiltonian nature of the twistor transformation between Minkowski
spaces ${\bf M}$ and ${\hat {\bf M}}$ \cite{pm}
is related to the invariance of the 1-form (\ref{typeIg})
\begin{equation}
\xi_{A} \, d \xi^{A}\, =  d \zeta
\label{inv1}
\end{equation}
which implies that the symplectic $2$-form
\begin{equation}
\omega = d \, \xi_{A} \, \wedge \, d \xi^{A}
\label{inv2}
\end{equation}
is also invariant under identification with warp.

\section{Holomorphic transformation}

So far we have dealt with conditions of Penrose's Type $I$ geometry
which insures continuity of the induced degenerate metric on ${\cal N}$.
This is a requirement restricted to the surface $v = 0$.
But the construction of the continuous metric on the full space
${\cal M} = {\bf M} \cup {\cal N} \cup {\hat {\bf M}}$ that describes
impulsive spherical gravitational waves requires information in addition
to eqs.(\ref{warp}). Now we must demand that the conditions for
Penrose's Type $II$ and $III$ geometry must be satisfied
as well. We start with continuity of the second fundamental form.
For this purpose we shall introduce another
$2$-component spinor ${\mu}^A$ with
flagpole direction along a straight null-path. Then the
$2$-component spinor ${\xi}^A$ will have flagpole direction
along a null cone with vertex on the straight null-path. For
the position vector of a general point we have therefore
\begin{eqnarray}
x^{\mu} \leftrightarrow x^{AX'} & = &
u\,\xi^A  \xi^{X'} + v\,\mu^A \mu^{X'} \nonumber\\
& = &      \begin{array}{c}
u\left(  \begin{array}{cc}
           1 & {\bar \zeta} \\
      \zeta  & \zeta  {\bar \zeta} \end{array} \right) +
v \left(  \begin{array}{cc}
0       &   0 \\
0       &   1
\end{array} \right)        \end{array}
\label{spinmat1}
\end{eqnarray}
from which one can read off the constant $2$-component spinor
\begin{equation}
 \mu^A=
 \left(  \begin{array}{c}
0  \\
1 \end{array} \right)
\label{cspinor}
\end{equation}
satisfying the normalization condition
\begin{equation}
{\xi}_A {\mu}^A = 1.
\label{normal}
\end{equation}
Now performing an arbitrary holomorphic transformation
$\zeta \rightarrow h(\zeta) $ we can present
the position vector of a general point in the form
\begin{equation}  \begin{array}{c}
\hat x^{AX'} =u|\eta|\left(  \begin{array}{cc}
1 & {\bar h} \\
h &   |h|^2     \end{array} \right) +
              v |\tilde \eta| \left(  \begin{array}{cc}
1       &   \bar m \\
m       &    |m|^2
\end{array} \right)     \end{array}
\label{spinmat2}
\end{equation}
and the $2$-component spinor $\mu^A$ can be written in a form
analoguous to eqs.(\ref{hspinor})
\begin{equation}
\hat \mu^A =
\left(  \begin{array}{c}
{\tilde \eta}^{1/2}  \\
m \,  {\tilde \eta}^{1/2} \end{array} \right)
\label{mspinor}
\end{equation}
where the two unknown holomorphic functions $m$ and ${\tilde \eta}$
have to be determined according to the requirements of Penrose's
Type $II$ and $III$ geometry. That is, the normalization condition
(\ref{normal}) and
\begin{equation}
{\mu}_A  \, \frac{d {\xi}^A }{d \zeta} = 0 =
{\hat \mu}_A  \, \frac{d {\hat \xi}^A }{d \zeta}
\label{contin}
\end{equation}
must be invariant under the holomorphic warp.
From eqs.(\ref{normal}) and (\ref{contin}) we find
\begin{equation}
m = h - 2 \,  \frac{ {h'}^{\,2} }{h''}
\label{mh}
\end{equation}
and
\begin{equation}
      {\tilde \eta} = \frac{h'}{( m - h )^2}
\label{etatilde}
\end{equation}
which completes the determination of the $2$-component spinor $\hat \mu^A$.
Substituting the expressions (\ref{etatr}), (\ref{mh}) and
(\ref{etatilde}) into eq.(\ref{spinmat2}) we find the
transformation to the flat region $v>0$ behind the impulsive wave
\begin{eqnarray}
U'&=&\frac{u}{ |h'|} + \frac{v}{4|h'|} \left|\frac{h''}{h'}\right|^2,
  			\nonumber \\
V'&=&\frac{u}{ |h'|}\, |h|^2 +
\frac{v}{4|h'|} \left|\frac{h''}{h'}\right|^2 |m|^2, \label{wtrans}\\
Z'&=&\frac{u}{ |h'|}\, h + \frac{v}{4|h'|} \left|\frac{h''}{h'}\right|^2 m
\nonumber
\end{eqnarray}
which highlights the role played by the spinors ${\xi}^A$ and ${\mu}^A$.
With this transformation the metric (\ref{harika}) is obtained readily.
It is a continuous metric for all values of $v$ which is a Petrov Type $N$
exact solution Einstein's vacuum field equations.

\subsection{Properties}

    It will be useful to recall some properties of the Schwarzian
derivative \cite{yoshida} because they play an important role in the
physical interpretation of the metric (\ref{harika}) as the radiation
accompanying the snapping of a cosmic string.
First of all we have $PGL(2,C)$-invariance
\begin{equation}
   \left\{ \frac{ \alpha h + \beta}{\gamma h +\delta} \, ; \zeta \right\} =
   \{ h \, ; \zeta \} \;\;\;\;\;\;
\forall  \; \left( \begin{array}{cc}
    \alpha & \beta \\ \gamma & \delta    \end{array}
    \right) \in GL(2,C)
\label{pgl2c}
\end{equation}
which shows that the metric (\ref{harika}) is one of the simplest
examples in twistor theory.
Using the connection formula where under an arbitrary analytic
change $\zeta \rightarrow z$ we have
\begin{equation}
   \{ h \, ; z \} \, d z^2 =   \{ \zeta ; z \} \, d z^2 \; + \;
   \{ h \, ; \zeta \} \, d \zeta^2  \, ,
\label{connfor}
\end{equation}
that is, $h(\zeta)$ is $PGL(2,C)$-multi-valued if any of its two branches
are projectively related, but $\{ h \, ; \zeta \}$ is single-valued.
It is the single-valued Schwarzian derivative that appears in the metric.
Further, if we consider the null co-frame
\begin{equation}
     l=dv,\;\;\; n=du, \;\;\; m= u \, d {\bar W}_1 + v \theta (v) \, d W_2
\end{equation}
where $d W_1$ and $ d W_2$ are Abelian differentials,
then the condition for the quadratic form $ d W_1 \, d W_2$
to be Schwarz-integrable is
\begin{equation}
 d W_1 \, d W_2   = \{ h ; \zeta \} \, d \zeta^2
\end{equation}
which is precisely the form realized in the metric (\ref{harika}).

The phase of a spinor is defined by the stereographic projection of
the intersection of the null cone with the unit sphere. It follows
from eq.(\ref{2spinor}) that it is the phase that undergoes the warp.
If the warp were to be given by a M\"obius transformation
\begin{equation}
  h = \frac{ \alpha \zeta + \beta}{\gamma \zeta +\delta}
\label{mobius}
\end{equation}
which is simply a uniform rotation of the Riemann sphere, then it would
not be a true warp at all and there will be no impulsive spherical
gravitational wave. This is confirmed by the vanishing of
the Schwarzian derivative for $h$ given by eq.(\ref{mobius}).

   The relationship between the impulsive spherical gravitational wave
and snapping cosmic string is related to the local behaviour
of the Schwarzian. We shall be interested in the choice of the warp function
\begin{equation}
  h = \left( \frac{ P \,\zeta - Q }{ R \,\zeta - S }   \right)^s
\label{warpstr}
\end{equation}
and without loss of generality we can take $P=1, Q=R=S=0$ so that we can
simply consider
\begin{equation}
h = \zeta^s
\label{local}
\end{equation}
where $s$ is a constant different from one. Note that here we could
have considered a more general form $ h = \zeta^s f(\zeta) $ where $f$
is holomorphic, or $\ln|\zeta f(\zeta)|$ which is its $``s=0"$ form.
But this is an inessential refinement and we shall not consider it any
further. From the choice of the warp function (\ref{local}) we get
\begin{equation}
 \{ h \, ; \zeta \} \, = - \frac{1 - s^2}{4 \zeta^2}
\label{localres}
\end{equation}
thus in the metric we have a meromorphic differential with a double pole.
In this paper we shall restrict our discussion
to real values of the parameter $s$. To exhibit the cosmic string aspect
of the metric (\ref{harika}) with (\ref{localres})
we transform to new coordinates
\begin{eqnarray}
u&=&\frac{1}{4}\, \left[(s+1)^2 \,\rho^{s-1} \,U'
- (s-1)^2 \,\rho^{-1-s} \,V'\right],  \nonumber \\[1mm]
v&=& \rho^{1-s} \left( V'- \rho^{2 s} \,U'\right), \label{cstrans}\\[1mm]
\zeta&=&\rho\, e^{i \phi}
\nonumber
\end{eqnarray}
where we have introduced the definitions
\begin{eqnarray}
\rho = \left[\frac{R+\sqrt{R^2+2 (s^2-1)\, U'V'}}{\sqrt{2}\,(s+1) \,U'}
\right]^{1/s},  R=(Z' \bar{Z'})^{1/2} \label{rho} \\
 U'=\frac{t-z}{\sqrt{2}},\qquad \qquad  V'=\frac{t+z}{\sqrt{2}}. \nonumber
\end{eqnarray}
This brings the metric given by (\ref{harika}) and (\ref{localres})
to the form
\begin{equation}
ds^2 = dt^2-dz^2 -dR^2 - s^2 R^2 d\phi^2
\label{csmetric}
\end{equation}
which is flat space with a conical deficit determined by the parameter $s$.
It is the metric around a straight cosmic string.

  Physical properties of the metric (\ref{harika}) can now be summarized.
Singularities of the solution (\ref{harika}) which are due to the
necessarily-incomplete spherical wavefronts are related to conical
singularities of cosmic strings as we saw above. The choice
of the warp function (\ref{local}) where $s$ is a real constant
different from one results in the flat metric (\ref{csmetric}) with
conical deficit for $v>0$. For $v<0$ we have complete Minkowski space.
At $v=0$ there is an impulsive spherical wave. This is the scenario for
the creation of a cosmic string with an impulsive spherical wave
at its ends. The opposite scenario is physically more relevant.
We start with a cosmic string, Minkowski space with conical deficit,
and the string snaps at $v=0$ with the emission of an impulsive
spherical wave and Minkowski space without conical singularity
is left behind. Currently acceptable values for the mass
per unit length $\mu$ of a cosmic string suggest
$$ s = 1 + \epsilon \, \;\;\;\;\; \epsilon = G \mu \approx 10^{-6}$$
and the formulae (\ref{cstrans}) and (\ref{rho}) for the transformation of
coordinates to Minkowski space determine the deficit angle
of the conical singularity in terms of $s$.

   In the discussion of the cosmic string above we took the constant $s$
to be real. But there is no reason why it cannot be complex. In this
case the imaginary part of $s$ will impart rotation to the cosmic string
and we get the metric of a spinning cosmic string which is
Lorentz-invariant and does not violate
causality, in contrast to an alternative suggestion for spinning string
due to Deser, Jackiw and 't Hooft \cite{djt}. This is also an important
issue which will be discussed in a seperate publication \cite{an1}.

  Finally, it will be useful to compare the continuous metric for an
impulsive $pp-$wave
\begin{equation}
  d s^2 = 2 \, d \, u \; d \, v  -
   2 \left| d \, {\bar \zeta} \; +
      v \, \theta(v) \, q_{\zeta \zeta}
    d \, \zeta \right|^2
\label{pp}
\end{equation}
where $q$ is an analytic function
\begin{equation}
 q_{\zeta {\bar \zeta} }    = 0
\label{ppeq}
\end{equation}
with that of the impulsive spherical wave given in eqs.(\ref{harika}) and
(\ref{ppeq}).
The impulsive $pp-$waves are obtained in limit $\lambda \rightarrow 0$
where \cite{rt1}
\begin{equation}   \begin{array}{cll}
u & \rightarrow & \lambda^{-1} u + \lambda^{-2}, \\
v & \rightarrow & \lambda v, \\
\zeta & \rightarrow & \lambda^{2} \zeta, \\
\{\, h \, ; \zeta \, \} & \rightarrow & \lambda^{-3}  \; q_{\zeta \zeta}
\end{array}   \label{limit}
\end{equation}
so that we have the following correspondance
\begin{equation}   \begin{array}{cll}
h = \zeta^s & \;\;\; \rightarrow \;\;\; & q =
\ln \zeta {\bar \zeta} \, ,  \\
h = e^{s \zeta} & \;\;\; \rightarrow \;\;\; & q = \frac{1}{2}
\left( \zeta^2 + {\bar \zeta}^2 \right)
\end{array}   \label{corresp}
\end{equation}
whereby the metric of the snapping cosmic string goes over into
the Aichel\-burg-Sexl solution \cite{as} and the Gleiser-Pullin
solution \cite{gp} corresponds to the impulsive plane wave \cite{rp1}.

   There is, however, one important respect in which the $pp$-wave
metric is different from the metric for spherical waves.
Namely, in the case of shock waves where the
Riemann tensor suffers a Heaviside step
function discontinuity rather than the $\delta$-function discontinuity
characteristic of impulsive waves, it is sufficient \cite{szekeres}
to replace $v \theta (v)$ by $v^2 \theta (v)$ in the impulsive $pp$-wave
metric of eq.(\ref{pp}). If we were to try the same procedure
for spherical waves using eq.(\ref{harika}), we would find that
the resulting metric fails to be
an exact solution of vacuum Einstein field equations. The case of
spherical shock waves requires a different treatment which has been
given in \cite{nutku}.

\section{Non-null solutions}

   We have seen that the metric (\ref{harika}) describing
an impulsive spherical gravitational wave is constructed by Penrose's
identification with warp of two halves of Minkowski space with metric
given by eq.(\ref{cone1}) where $v=0$ is a null cone. There should
be a family of exact solutions of the Einstein field equations describing
impulsive spherical waves of which eq.(\ref{harika}) is the simplest
example. Such solutions will contain new parameters and the question
naturally arises as to the existence of a general technique for finding
them. The answer \cite{nutku} is simple. We must look for metrics describing
Minkowski space where $v=0$ is again a null cone but the metric admits
a set of new parameters.

   The simplest such metric is obtained by transforming the coordinates
in eq.(\ref{minkowski}) by
\begin{eqnarray}
  U'&=&\frac{k}{2}\,v  + \frac{u}{p},   \nonumber \\
  V'&=& v + \frac{u}{p}\,|\zeta|^2,         \label{ptrans}\\
  Z'&=&\frac{u}{p}\,\zeta                \nonumber
\end{eqnarray}
where
\begin{equation}
  p= 1 + \frac{k}{2}\,|\zeta|^2, \qquad   k = 0,\;\pm1
\label{parameters}
\end{equation}
and we obtain the metric
\begin{equation}
  ds^{2} =  2\,d\,u \;d\,v + k \, d\,v^2
  -  2\frac{u^{2}}{p^2} \; d\,\zeta \;d\,\bar{\zeta}
\label{cone11}
\end{equation}
where $v$ is again a null coordinate and the hypersurface ${\cal N}$ given
by $v=0$ is also a null cone. We shall now show that the generalization of
the metric (\ref{harika}) which gives the impulsive spherical wave
solution including the arbitrary constant $k$ in the Robinson-Trautman
solutions is obtained by Penrose's identification with warp on both sides
of ${\cal N}$ for the metric (\ref{cone11}). This will also prepare the
ground for the inclusion of the acceleration parameter that we shall
present in the next section.

   We shall again use spinor techniques and as it is evident from
eqs.(\ref{ptrans}) in the non-null case the position vector of a
general point on the future null cone is given by
\begin{equation}
 x^{\mu} \leftrightarrow x^{AX'} = u\,\xi^A  \xi^{X'} + v\,\mu^{AX'}
\label{gvector}
\end{equation}
where we have introduced a constant second-rank spinor $\mu^{AX'}$
defined along the world-path which cannot be expressed as a bi-spinor.
The vector equivalent of $\mu^{AX'}$ will have magnitude proportional to
$k$ which will make it timelike, or spacelike for $k=+1, k=-1$ respectively.
The explicit form of the transformation (\ref{gvector}) is given by
\begin{equation}  \begin{array}{c}
x^{AX'} =\frac{\textstyle u}{\textstyle p}\left(  \begin{array}{cc}
1 & {\bar \zeta} \\
\zeta  & \zeta  {\bar \zeta} \end{array} \right) +
v \left(  \begin{array}{cc}
\frac{1}{2} k  &   0 \\
0       &    1
\end{array} \right)        \end{array}
\label{gspinmat1}
\end{equation}
so that the $2$-component spinor ${\xi}^A$ retains flagpole direction
along the future null cone with vertex on the world-path which, however,
is no longer null. Comparing eqs.(\ref{gvector}) and (\ref{gspinmat1})
we find that the normalization conditions
\begin{equation}
\mu_{AX'} \xi^A \bar{\xi}^{X'} = 1
\label{gnormal1}
\end{equation}
and
\begin{equation}
\mu_{AX'} \mu^{AX'}  = k
\label{gnormal2}
\end{equation}
are satisfied.

  Penrose's identification with warp is given by the arbitrary
holomorphic transformation $\zeta \rightarrow h(\zeta)$. It results
in the following expression for the position vector of a general point
\begin{equation}  \begin{array}{c}
\hat x^{AX'} =\frac{\textstyle u}{\textstyle {\cal P}}\, |\eta|
\left(  \begin{array}{cc}
1 & {\bar h} \\
h & |h|^2                         \end{array} \right) +
v  \left(  \begin{array}{cc}
\hat \mu^{00'}       &  \hat \mu^{01'} \\
\hat \mu^{10'}       &   \hat \mu^{11'}
\end{array} \right)     \end{array}
\label{gspinmat2}
\end{equation}
where
$$ {\cal P}=1 + \frac{1}{2} k \,|h|^2 $$
and we need to determine the spinors $\xi^A$ and $\mu^{A X'}$.
The $2$-component spinor $\xi^A$ has an expression analoguous to
eq.(\ref{2spinor})
\begin{equation}
 \hat \xi^A=
\left(  \begin{array}{c}
(\eta/{\cal P})^{1/2}  \\[3mm]
(\eta/{\cal P})^{1/2}\;h  \end{array} \right)
\label{ghspinor}
\end{equation}
and the requirement of Penrose's Type $I$ geometry, namely the
invariance of the $1$-form $\xi_{A} \, d \xi^{A}$
under a holomorphic mapping on the future null cone ${\cal N}$ leads to
\begin{equation}
|\eta|= \frac{1}{|h'|}\,\frac{1 + \frac{\textstyle k}{\textstyle 2}\,|h|^2}
{1 + \frac{\textstyle k}{\textstyle 2}\,|\zeta|^2}
\label{geta}
\end{equation}
as in eq.(\ref{typeIg}). In terms of coordinates we have
\begin{eqnarray}
 {\hat v} &=&0 \; = \; v,          \nonumber   \\[1mm]
{\hat u} &=& \frac{u}{|h'|}\,
\frac{1 + \frac{\textstyle k}{\textstyle 2}\,|h|^2}
{1 + \frac{\textstyle k}{\textstyle 2}\,|\zeta|^2},  \label{gwarp} \\[1mm]
  {\hat \zeta} &=& h\,(\zeta)     \nonumber
\end{eqnarray}
for the identification of two halves of Minkowski space given by
the metric (\ref{cone11}) along the null cone $v=0$.
In eqs.(\ref{geta}) and (\ref{gwarp}) we find the first indication
that will be repeated throught the subject. In identification with
warp $\zeta \rightarrow h(\zeta)$ is not to be taken as a mechanical
dictum. It is the continuity conditions along the null cone, in this case
the invariance of $\xi_A \, d \xi^A$, that are important and as we
find in these equations both $\zeta$ and $h(\zeta)$ will appear
in the final expression for the metric.

The explicit form of the second-rank spinor $\hat \mu^{AX'}$ is
obtained from Penrose's Type $II$ and $III$ geometry that
a holomorphic mapping with warp must preserve the normalization
conditions (\ref{gnormal1}) and (\ref{gnormal2}) and
\begin{equation}
\mu_{AX'} \, \frac{d}{d \zeta} \,(\xi^A \bar \xi^{X'}) = 0
\label{gcontin}
\end{equation}
together with its complex conjugate that replace eqs.(\ref{contin}).
The solution of eqs.(\ref{gnormal1}), (\ref{gnormal2}) and (\ref{gcontin})
gives us
\begin{eqnarray}
\hat \mu^{00'}&=&\frac{1}{4 |h'|}\,\left(k\,K
+  p\, \left|\frac{h''}{h'}\right|^2 \right),        \nonumber \\[2mm]
\hat \mu^{11'}&=&\frac{1}{4 |h'|}\,\left[ k\;(|h|^2\,K -
2 \,h \,\bar h' \,\bar \zeta - 2 \,\bar h\, h' \, \zeta)
+  p\,|m|^2\, \left|\frac{h''}{h'}\right|^2 \right], \label{2ranksp} \\[2mm]
\hat \mu^{10'}&=&\frac{1}{4 |h'|}\,\left[ k\;(h\,K - 2 \, h' \,\zeta)
+  p\,m\, \left|\frac{h''}{h'}\right|^2 \right] \nonumber \\[2mm]
\hat \mu^{01'}&=&\overline{\hat \mu^{10'}},       \nonumber
\end{eqnarray}
where
$$K= 2 + \frac{h''}{h'}\,\zeta + \frac{\bar h''}{\bar h'}\,\bar \zeta$$
and $m$ is again given by eq.(\ref{mh}).
From eqs.(\ref{gspinmat2}) we obtain the generalization
of the transformation (\ref{wtrans}) for arbitrary values of
the parameter $k$
\begin{eqnarray}
V'&=&\frac{u}{p}\,\frac{|h|^2}{|h'|} +
\frac{v}{4 |h'|}\,\left[ k\;(|h|^2\,K -
2 \,h \,\bar h' \,\bar \zeta - 2 \,\bar h\, h' \, \zeta)
+  p\,|m|^2\, \left|\frac{h''}{h'}\right|^2 \right], \nonumber \\[2mm]
U'&=&\frac{u}{p}\,\frac{1}{|h'|} + \frac{v}{4 |h'|}\,\left(k\,K
+  p\, \left|\frac{h''}{h'}\right|^2 \right), \label{gwtrans} \\[2mm]
Z'&=&\frac{u}{p}\,\frac{h}{|h'|} + \frac{v}{4 |h'|}\,\left[ k\;(h\,K -
2 \, h' \,\zeta) +  p\,m\, \left|\frac{h''}{h'}\right|^2 \right] \nonumber
\end{eqnarray}
which for $k=+1$ is equivalent to the one given by Hogan \cite{hogan1}
but once again spinor methods bring simplicity and compactness.
From eqs.(\ref{gwtrans}) we arrive at the metric for an impulsive spherical
gravitational wave for arbitrary values of the parameter $k$ \cite{hogan1}
\begin{equation}
 d s^2 = 2\,d \,u \,d \, v +k\,d\,v^2 -
 2 \left|\frac{u}{p} \, d \, \bar \zeta +
 p\, v \, \theta(v) \{ h \, ; \zeta \}\,d \,\zeta \right|^2
\label{gharika}
\end{equation}
which is continuous. It shares
all the properties of the metric (\ref{harika}) and reduces to it
in the null case $k=0$. We note that, as we remarked earlier,
besides $h(\zeta)$ also $\zeta$ enters into the final result
for the metric through $p$ given by eq.(\ref{parameters}).

\section{Accelerating solution}

  We have shown that metrics generalizing the impulsive spherical
gravitational wave metric (\ref{harika}) are obtained by extending
Penrose's identification with warp to two halves of Minkowski space
where $v=0$ is a null cone but the Minkowski metric contains new parameters.
It was pointed out in \cite{nutku} that flat metric with constant
acceleration \cite{kw} is another such example. The metric for flat space
with constant acceleration is obtained by the transformation
\begin{eqnarray}
  V'&=& \frac{1}{a} \left(1 - e^{\;\textstyle-a v}\right)
  +  \frac{u}{p}\;e^{\;\textstyle-a v}\;|\zeta|^2,  \nonumber \\
  U'&=&\frac{k}{2\,a}\, \left(e^{\;\textstyle a v} -1 \right)
  + \frac{u}{p}\;e^{\;\textstyle a v},              \label{acctrans} \\
  Z'&=&\frac{u}{p}\,\zeta ,  \nonumber
\end{eqnarray}
where $a$ is the acceleration parameter and $p$ is given by
(\ref{parameters}). Applying the transformation (\ref{acctrans}) to
the metric (\ref{minkowski}) we obtain
\begin{eqnarray}
 d s^2 & = & 2\,d \,u \,d \, v +\left(k +
 \frac{\textstyle 2\, a\,u}{\textstyle p}\right)
 \left(1 - \frac{\textstyle a\,u}{\textstyle p}\,|\zeta|^2 \right)\,d\,v^2
   \nonumber \label{accmink}\\ & &
+ \frac{\textstyle 2\, a\,u^2}{\textstyle p^2}\,(\bar \zeta\, d \,\zeta
 + \zeta\, d \,\bar \zeta)\,d\,v
- 2 \,\frac{\textstyle u^2}{\textstyle p^2} \, d \,\zeta \, d \,\bar \zeta
\end{eqnarray}
which is the flat-space limit of the uniformly accerelating $C$-metric
\cite{kw} in stereographic coordinates on the sphere
and reduces to the metric (\ref{cone11}) for $a=0$. For
the spinor description of the metric (\ref{accmink})
it is useful to introduce new coordinates
\begin{eqnarray}
V &=& \frac{1}{a} \left(1 - e^{\;\textstyle-a v}\right),   \nonumber \\
U &= & u\,e^{\;\textstyle a v},                     \label{newcoord} \\
Z & =& \zeta\, e^{\;\textstyle-a v}                  \nonumber
\end{eqnarray}
and the transformation (\ref{acctrans}) becomes
\begin{eqnarray}
  V'&=& V + \frac{U}{p}\;|Z|^2 			\nonumber \\
  U'&=&\frac{k}{2}\,\frac{V}{1-a\,V}  + \frac{U}{p}  \label{newtrans} \\
  Z'&=&\frac{U}{p}\,Z ,  \nonumber
\end{eqnarray}
where $p$ is now given by
\begin{equation}
p = 1+ \frac{\textstyle k}{\textstyle 2 \,(1-a\,V)^2} \,|Z|^2
\label{newp}
\end{equation}
which depends on the acceleration parameter as well.
In these coordinates the metric (\ref{accmink}) assumes the form
\begin{eqnarray}
 d s^2 & = & 2 \,d \, V \left[ d\,U +
 \frac{\textstyle k}{\textstyle 2}\,
\frac{\textstyle d\,V}{\textstyle (1-a\,V)^2}+
\frac{\textstyle 2\, (1-p)}{\textstyle p} \,
\frac{\textstyle a\,U}{\textstyle 1-a\,V}\,d\,V \right]
\nonumber     \\[2mm] & &
- \frac{\textstyle 2\,U^2}{\textstyle p^2}\, d \,Z \, d \,\bar Z
\label{accmink2}
\end{eqnarray}
which has a removable singularity at $V = a^{-1}$.
The hypersurface $V=0$ is again a null cone and we can construct a new
impulsive spherical wave metric by identifying with warp the two
halves Minkowski space defined by $V>0$ and $V<0$ in the metric
(\ref{accmink2}).

   The position vector of a general point on the null cone of
a uniformly accelerating world-path has the form
\begin{eqnarray}
x^{AX'} & = & U\,\xi^A  \xi^{X'} + V\,\mu^{AX'} \label{accvector}\\
&=&  \begin{array}{c}
\frac{\textstyle U}{\textstyle p}\,
\left(  \begin{array}{cc}
1 & {\bar Z} \\
Z  & |Z|^2 \end{array} \right) +
V \left(  \begin{array}{cc}
\frac{\textstyle k}{\textstyle 2\,(1-a\,V)  }        &   0 \\
0       &    1
\end{array} \right)     \end{array}
\label{accspinmat1}
\end{eqnarray}
which follows from eqs.(\ref{newtrans}). The tangent vector to the path
is given by the second rank spinor $\mu^{AX'}$.
From eqs.(\ref{accvector}) and (\ref{accspinmat1}) we find that the
spinors $\xi^A$ and $\mu^{AX'}$ satisfy the normalization  conditions
\begin{equation}
\mu_{AX'} \xi^A \bar{\xi}^{X'} = F
\label{accnormal1}
\end{equation}
where
\begin{equation}
F=\frac{1}{p}\left(1+ \frac{\textstyle k}{\textstyle 2\,(1-a\,V)} \,|Z|^2
\right)
\label{norfunc}
\end{equation}
and
\begin{equation}
\mu_{AX'} \mu^{AX'}  = \frac{\textstyle k}{\textstyle 1-a\,V}.
\label{accnormal2}
\end{equation}
The warp will now be given by $Z\rightarrow h(Z)$, however, we must
repeat our earlier remark that it must not be taken as a dictum.
In particular, the normalization condition (\ref{accnormal1})
involves $\zeta$ through $F$, unlike the situations encountered earlier.
It must remain unchanged under the warp.

     Under the arbitrary holomorphic transformation
of the spin-frame the position vector of a general point is given by
\begin{equation}  \begin{array}{c}
\hat x^{AX'} =\frac{\textstyle U}{\textstyle P}\, |\eta|
\left(  \begin{array}{cc}
1 & {\bar h } \\
h & |h|^2 \end{array} \right) +
V \left(  \begin{array}{cc}
\hat \mu^{00'}       &  \hat \mu^{01'} \\
\hat \mu^{10'}       &   \hat \mu^{11'}
\end{array} \right)     \end{array}
\label{accspinmat2}
\end{equation}
where
$$ P = 1+ \frac{\textstyle k}{\textstyle 2 \,(1-a\,V)^2}\,|h|^2. $$
It follows that we may now introduce the $2$-component spinor
\begin{equation}
 \hat \xi^A=
\left(  \begin{array}{c}
(\eta/P)^{1/2}  \\[3mm]
(\eta/P)^{1/2}\;h  \end{array} \right)
\label{acchspinor}
\end{equation}
where
\begin{equation}
|\eta|= \frac{1}{|h'|}\,\frac{P}{p}
\label{acceta}
\end{equation}
which insures that $\,\xi_{A} \, d \xi^{A}$
remains invariant under the arbitrary holomorphic transformation.
Thus the identification with warp of the two halves of Minkowski space
with metric (\ref{accmink2}) across $V =0 $ is given by
\begin{eqnarray}
 {\hat V} &=&0 \; = \; V \nonumber  \\[1mm]
  {\hat Z} &=& h\,(Z) \label{accwarp}  \\[1mm]
  {\hat U} &=& \frac{U}{|h'|}\,\frac{1 + \frac{\textstyle k}{\textstyle 2}\,|h|^2}
{1 + \frac{\textstyle k}{\textstyle 2}\,|Z|^2}\; . \nonumber
\end{eqnarray}
The explicit expression for the components of the second-rank spinor
$\hat \mu^{AX'}$ follows from the requirement of Penrose's
Type $II$ and $III$-geometry \cite{bible}. The second fundamental form
of ${\cal N}$ must be the same when induced by the two flat pieces $(V<0)$
and $(V>0)$ of Minkowski space. This implies the invariance of the
normalization conditions (\ref{accnormal1}) and (\ref{accnormal2}).
From the solution of these equations we arrive at the
transformation of coordinates in the region $V>0$
\begin{eqnarray}
V'&=&\frac{U}{p}\,\frac{|h|^2}{|h'|}
+ \frac{\textstyle V}{4 |h'|}\,\left[
 q\,|m|^2\, \left|\frac{h''}{h'}\right|^2 +
 \right. \nonumber \\ [2mm] & & \left.
 \mbox{ }
  \frac{\textstyle k}{1 -a\,V}\,
|h|^2\,\left(2 + \frac{h''}{h'}\,\frac{m}{h}\,Z
+ \frac{\bar h''}{\bar h'}\,\frac{\bar m}{\bar h}\,\bar Z \right) \right]
 			\nonumber \\[2mm]
U'&=&\frac{U}{p}\,\frac{1}{|h'|}
+\frac{\textstyle V}{4 |h'|}\,\left[ q\, \left|\frac{h''}{h'}\right|^2
+ \frac{\textstyle k}{1 -a\,V}\,
\left(2 + \frac{h''}{h'}\,Z + \frac{\bar h''}{\bar h'}\,\bar Z \right)
 \right] \label{accwtrans} \\ [2mm]
Z'&=&\frac{U}{p}\,\frac{h}{|h'|}
+ \frac{\textstyle V}{4 |h'|}\,\left[
 q\,m\, \left|\frac{h''}{h'}\right|^2 +
\right. \nonumber \\ [2mm] & & \left.
 \mbox{ }
 \frac{\textstyle k}{1 -a\,V}\,
h\,\left(2 + \frac{h''}{h'}\,\frac{m}{h}\,Z
+ \frac{\bar h''}{\bar h'}\,\bar Z \right) \right] \nonumber
\end{eqnarray}
where we have introduced
\begin{equation}
q = 1+ \frac{\textstyle k}{2(1 -a\,V)}\,|Z|^2
\label{q}
\end{equation}
and $m$ is again given by (\ref{mh}).

    The continuous impulsive spherical gravitational wave
metric obtained by letting the vertex of the null cone to
lie on a world-line with constant acceleration is given by
\begin{eqnarray}
 d s^2 & = & 2 \,d \, V \left[ d\,U +  \frac{\textstyle k}{\textstyle 2}
 \,\,\frac{\textstyle d\,V}{\textstyle (1-a\,V)^2}
+ \frac{\textstyle 2\, (1-p)}{\textstyle p} \,
\frac{\textstyle a\,U}{\textstyle 1-a\,V}\,d\,V
 \right. \nonumber \\ [2mm] & & \left.
 \mbox{ }
+\, \frac{\textstyle k\,a}{\textstyle {2\,(1- a\,V)^2}}\,\,
V^2 \,\theta(V)
\left[\{ h \, ; Z \}\,Z\,d \,Z +
\{ \bar h \, ; \bar Z \}\,\bar Z\,d \,\bar Z \,\right]\,\right]
\label{accmetric1}  \\ [2mm] & &
- 2 \,\left|\frac{U}{p} \,d \, \bar Z +
q\,V \,\theta(V)\, \{ h \, ; Z \}\,d \,Z \right|^2.  \nonumber
\end{eqnarray}
It follows from the non-linear holomorphic transformation (\ref{accwtrans})
of Minkowski spacetime.
In succesive limits $a=0$ and $k=0$ it reduces to (\ref{gharika})
and (\ref{harika}), whereas for $k=0$ it immediately reduces to the null case
(\ref{harika}) because acceleration is not meaningful
when the world-path is null.

\section{Curvature}

   In order to show that the metric (\ref{accmetric1}) satisfies
the Einstein vacuum field equations and describes an impulsive spherical
gravitational wave we shall now calculate its curvature using the
Newman-Penrose formalism \cite{newmp}. We start with the null co-frame
\begin{eqnarray}
l & = & d\,V,            \nonumber \\ [2mm]
n & = & d\,U + \left[\frac{\textstyle 2\,a\,U (1-p)}{\textstyle p} \,
+ \frac{\textstyle k}{2}\,\frac{1}{\textstyle 1-a\,V}\right]
\,\frac{d\,V}{\textstyle 1-a\,V}
\nonumber \\ [2mm] & &
+\, \frac{\textstyle k\,a}{\textstyle 2}\,\,
\frac{V^2 \,\theta(V)}{\textstyle {(1- a\,V)^2}}\,\,
\left[\{ h \, ; Z \}\,Z\,d \,Z +
\{ \bar h \, ; \bar Z \}\,\bar Z\,d \,\bar Z \,\right]\,d\,V,
\label{coframe} \\
m & = & \frac{U}{p} \,d \, Z +
q\,V \,\theta(V)\, \{ \bar h \, ; \bar Z \}\,d \,\bar Z  \nonumber
\end{eqnarray}
and the metric (\ref{accmetric1}) is given by
\begin{equation}
 d s^2  =  l\otimes n + n\otimes l -  m\otimes \bar m - \bar m\otimes m
\label{accnull}
\end{equation}
the Newman-Penrose null form.
Calculating the spin-coefficients with the co-frame (\ref{coframe})
we obtain
\begin{eqnarray}
\kappa & = & \tau = \pi = \epsilon = 0,         \nonumber \\ [2mm]
\rho & = & - \frac{1}{\Delta}\,\frac{U}{p^2},  \nonumber   \\ [2mm]
\nu & = &  \frac{\textstyle k\,a}{\textstyle p}\,\,
\frac{\bar Z}{\textstyle {(1- a\,V)^3}},        \nonumber   \\ [2mm]
\sigma & = & \frac{1}{\Delta}\,\,\frac{q}{p}\,\,V\,\theta(V)\,
\{ \bar h \, ; \bar Z \},               \label{spincoeff}  \\  [2mm]
\gamma & = &  \frac{\textstyle {1-p}}{\textstyle p}\,\,
\frac{\textstyle a}{\textstyle {1- a\,V}},     \nonumber   \\ [2mm]
\lambda & = & \frac{\{ h \, ; Z \} }{\Delta}\, \theta(V)\,
\left[ U + \frac{\textstyle k}{\textstyle 2 p}\,\,
\frac{\textstyle q\,V}{\textstyle {(1-a\,V)^2}} \right], \nonumber   \\ [2mm]
\mu & = & - \frac{1}{\Delta} \left[\frac{\textstyle k}{\textstyle 2 p^2}\,\,
\frac{\textstyle U}{\textstyle {(1-a\,V)^2}}
+ p\,q\,V\,\theta(V)\,\{ h \, ; Z \}\,\{ \bar h \, ; \bar Z \} \right],
\nonumber   \\ [2mm]
\beta & = & -\bar \alpha = \frac{\textstyle k}{\textstyle {4 p\,\Delta}}\,\,
\frac{1}{\textstyle {(1-a\,V)^2}}\,\left[\frac{U}{p}\, Z +
 q\,V \,\theta(V)\, \{ \bar h \, ; \bar Z \}\,\bar Z \right]
    \nonumber
\end{eqnarray}
where
\begin{equation}
\Delta = \frac{U^2}{p^2} - q^2\,V^2\,\theta(V)\,
\{ h \, ; Z \}\,\{ \bar h \, ; \bar Z \}.
\label{delta}
\end{equation}
Using these spin coefficients we find that all Ricci scalars
vanish identically for all values of $V$, while in the curvature
the only nonvanishing Weyl tetrad scalar
\begin{equation}
\Psi_{4} = -\frac{\{ h \, ; Z \}}{U}\,p^2 \,\delta(V)
\label{curv}
\end{equation}
suffers a Dirac $\delta$-function discontinuity. Thus the metric
(\ref{accmetric1}) is of the Petrov type $N$ exact solution of the Einstein
vacuum field equations. It should be noted that the constant
acceleration parameter in the metric (\ref{accmetric1}) does not appear
in the curvature. This is a well-known fact that is repeated for the case
of uncharged metrics \cite{kw},\cite{pd}.

\section{Conclusion}

  The continuous form of impulsive gravitational wave metrics follows from
Penrose's identification with warp. For impulsive spherical waves Minkowski
space is cut into two pieces along a null cone which are then identified
with warp, an arbitrary holomorphic transformation. Using the $2$-component
spinor formalism we have presented the explicit form of the holomorphic
transformations that lead to the family of impulsive spherical
wave metrics. For waves with smooth profile the resulting metrics
are equivalent to Petrov Type $N$ Robinson-Trautman solutions.
In the continuous form of the metric for impulsive spherical waves
the ``wire singularities" in the Robinson-Trautman solutions reappear
as Minkowski space with and without a conical deficit outside
and inside the future null cone respectively. Therefore the physical
interpretation of these solutions is that a snapping cosmic string is the
source of the impulsive spherical wave. The full power of Penrose's method
of identification with warp becomes manifest with $2$-component spinor
techniques and leads to the simplest derivation of the exact solutions.
In this framework we have discussed the derivation known solutions and
then used these tools to construct a new exact solution of the Einstein
field equations that describes an impulsive spherical gravitational
wave in accelerating frame.

\end{document}